\documentclass[pdflatex,sn-nature]{sn-jnl}

\usepackage{graphicx}%
\usepackage{multirow}%
\usepackage{amsmath,amssymb,amsfonts}%
\usepackage{amsthm}%
\usepackage{mathrsfs}%
\usepackage[title]{appendix}%
\usepackage{xcolor}%
\usepackage{textcomp}%
\usepackage{manyfoot}%
\usepackage{booktabs}%
\usepackage{algorithm}%
\usepackage{algorithmicx}%
\usepackage{algpseudocode}%
\usepackage{listings}%

\theoremstyle{thmstyleone}%
\theoremstyle{thmstyletwo}%

\theoremstyle{thmstylethree}%

\begin{document}

\title[Article Title]{Dispersion Control of Chiral Exciton-Polariton Transport with Dielectric Metasurfaces}


\author[1]{\fnm{J.} \sur{Jeon}}\equalcont{These authors contributed equally to this work.}

\author[1,2,3]{\fnm{K.} \sur{Wang}} \equalcont{These authors contributed equally to this work.}

\author*[1,2,3,5]{\fnm{Y.-C.} \sur{Wei}}
\equalcont{These authors contributed equally to this work.}\email{yc.wei@mx.nthu.edu.tw}

\author[1,2,3]{\fnm{F.} \sur{Cussiol}}

\author[1,2,3]{\fnm{M.} \sur{Berghuis}}

\author[2]{\fnm{F.}\sur{Xu}} 

\author[4]{\fnm{S.} \sur{Murai}}

\author[2]{\fnm{E.W.}\sur{Meijer}} 

\author*[1,2,3]{\fnm{J.}\sur{G\'omez Rivas}} \email{j.gomez.rivas@tue.nl}

\affil[1]{\orgdiv{Department of Applied Physics and Science Education}, \orgname{Eindhoven University of Technology}, \orgaddress{\postcode{5600MB}, \city{Eindhoven}, \country{The Netherlands}}}

\affil[2]{\orgdiv{Institute for Complex Molecular Systems}, \orgname{Eindhoven University of Technology}, \city{Eindhoven}, \country{The Netherlands}}

\affil[3]{\orgdiv{Casimir Institute}, \orgname{Eindhoven University of Technology}, \city{Eindhoven}, \country{The Netherlands}}

\affil[4]{\orgdiv{Department of Physics and Electronics,} \orgname{Graduate School of Engineering,
Osaka Metropolitan University}, \orgaddress{\street{599-8531}, \city{Osaka}, \country{Japan}}}

\affil[5]{\orgdiv{Department of Chemistry,} \orgname{National Tsing Hua University},  \city{Hsinchu}, \country{Taiwan}}

\abstract{
Exciton–polaritons provide a powerful platform for manipulating hybrid light–matter states with low effective masses and strong nonlinearities. Introducing chirality into these quasiparticles enables selective control over their spin and propagation, opening new opportunities for chiral transport and spin-selective polaritonic devices. We exploit the strong chiral light–matter coupling in silicon metasurfaces composed of tilted nanorod dimers to demonstrate selective transport of organic chiral exciton–polaritons. The metasurface supports surface lattice resonances and quasi-bound states in the continuum that simultaneously provide high photonic confinement and extrinsic chirality, giving rise to chiral exciton–polaritons in the achiral molecules. These exciton-polaritons exhibit a large magnitude of the dissymmetry factor, reaching a value of 0.93. Using photoluminescence Fourier microscopy and real-space imaging, we show that chiral exciton-polaritons propagate over distances exceeding 50 $\mu$m without significant degradation of their dissymmetry, with characteristic propagation lengths of approximately 6-13 $\mu$m. These propagation lengths correspond to an enhancement of 3 orders of magnitude compared to bare excitons. This work constitutes the first demonstration of enhanced and selective chiral transport of organic exciton–polaritons, driven by strong light–matter coupling, in achiral metasurfaces, paving the way for spin-selective polaritonic technologies using simple metasurfaces.}

%




\maketitle

\section{Introduction}\label{sec1}

Exciton-polaritons (EPs) are hybrid quasiparticles that form when photons in an optical cavity strongly couple to excitons (bound electron–hole pairs) in a semiconductor inside the cavity~\cite{Weisbuch1992,Lidzey1998,Deng2010,Bellessa2004}.
EPs are very light, with effective masses $\sim10^{-4}$ to $10^{-5}$ times the electron mass, due to their photonic character. This property, together with the strong nonlinearities due to their excitonic character, makes EPs very interesting in various applications, such as low-threshold polariton lasing or Bose-Einstein condensation~\cite{Imamoglu1996,Kasprzak2006,Christopoulos2007,
Kna-Cohen2010,Plumhof2014,Daskalakis2014,Byrnes2014,Ramezani2017,Bloch2022}, all-optical switching~\cite{Baas2004,Amo2010,Ballarini2013,Zasedatelev2021,Chen2022}, and quantum simulation~\cite{amo2016exciton,Ballarini2020,ghosh2020quantum,Mirek2021,kavokin2022polariton,barrat2024qubit}. The strong binding energy of Frenkel excitons in organic semiconductors, resulting from the weak dielectric screening, makes them particularly relevant for strong light-matter coupling. This binding energy leads to large oscillator strengths, facilitating strong coupling~\cite{Lidzey1998,Bellessa2004}. Moreover, EPs can be formed at room temperature in organic systems as the exciton binding energy is larger than the thermal energy, $k_BT$, with $k_B$ the Boltzmann's constant and $T$ the temperature. As a result of the strong binding energy, excitons in organic semiconductors are confined to single molecules, propagating only a few nanometers by hopping to nearby molecules~\cite{Mikhnenko2015}. One of the most remarkable properties of EPs is their enhanced transport length, which is associated with their ballistic propagation and low effective mass~\cite{Freixanet2000,Zakharko18,Tichauer23}. Indeed, EPs with propagation lengths of tens to hundreds of microns have been demonstrated in organic systems~\cite{Coles2014,Lenario2017,Rozenman18,Hou2020,Berghuis2022,balasubrahmaniyam2023enhanced,Dang2024}. This property has led to a renewed interest in the field of polaritonics, i.e., the domain between photonics (manipulation of light) and electronics (manipulation of charge), where light–matter hybridization enables phenomena and device functionalities that are impossible with either photons or excitons alone~\cite{LIEW2011, Sanvitto2016}.

In the field of electronics, spintronics harnesses the electron spin degree of freedom to transmit, store, and process information~\cite{Wolf2001,Zutic2004,Manchon2015}. In addition, valleytronics exploits the valley degree of freedom of charge carriers in certain crystalline materials, such as 2D transition metal dichalcogenides, graphene, and topological materials, to control the propagation of electronic spins~\cite{Rycerz2007,Xiao2012,Schaibley2016,liu2020generation,Wurdack2026}. Valleytronics can be viewed as an extension of spintronics concepts to the valley degree of freedom in momentum space, which also lead to chiroptical responses. Another method of utilizing electron spin is to exploit the coupling between the electron linear momentum and spin orientation in chiral materials, which gives rise to the chiral-induced spin selectivity (CISS) effect. This phenomenon enables controllable spin transport and the generation of both pure spin currents and spin-polarized charge currents~\cite{bloom2024chiral,naaman2012chiral}. An analogy to these phenomena in polaritonics would be the study and utilization of the spin angular momentum (SAM) of EPs through the hybridization of excitons with surface chiral optical modes that have positive (right-handed circular polarization, RCP) or negative (left-handed circular polarization, LCP) photon spin along their propagation direction. Similarly, selective propagation of chiral EPs should be enabled by the crystal structure that supports the photon mode, which strongly couples to excitons. Polariton spin separation based on achiral photonic structures and chiral excitons~\cite{chervy2018room,shi2025coherent,chen2025polariton}, and topological polariton propagation~\cite{smirnova2024polaritonic} have been demonstrated recently, offering new perspectives for spin-polarized transport. 

In this manuscript, we demonstrate spin-selective transport of chiral EPs formed by excitons in thin films of achiral molecules strongly coupled to extrinsically chiral optical modes on dielectric metasurfaces. The metasurfaces are composed of arrays of tilted silicon bars that support surface lattice resonances (SLRs) and quasi-bound states in the continuum (quasi-BICs)~\cite{Zhu2025}. The organic film consists of a perylene dye [N, N'-Bis(2,6-diisopropylphenyl)-1,7- and -1,6-bis
(2,6-diisopropylphenoxy)-perylene-3,4:9,10-tetracarboximide]~\cite{patent}, blended at high concentration in polymethyl methacrylate (PMMA). The molecular structures are shown in Supplementary Information, Fig.~S1. This dye is highly photo-stable and does not aggregate at high concentrations (35\%), which is required for strong light-matter coupling~\cite{Ramezani2017}. The hybrid EPs in this system inherit the chiral properties of the optical modes of the metasurface, which exhibit different SAM along different in-plane wave vectors. This characteristic allows the separation of RCP-EPs and LCP-EPs by the enhanced EP transport in the strongly coupled system and the subsequent control of the transport of chiral EPs. The chiral EPs have propagation lengths on the order of several microns, reaching distances longer than 50 $\mu$m without loss of dissymmetry. These exceptional propagation lengths are three orders of magnitude longer than the propagation length of excitons in uncoupled systems.   Looking ahead, further developments of chiral polaritonic systems may open pathways to integrate spin-selective transport with functional optoelectronic devices, ultimately bridging fundamental chiral science with practical quantum technologies.

\section{Results}\label{results}

\subsection{Extrinsically Chiral Metasurfaces}
\begin{figure}[!h]
\centering
\includegraphics[width=0.9\textwidth]{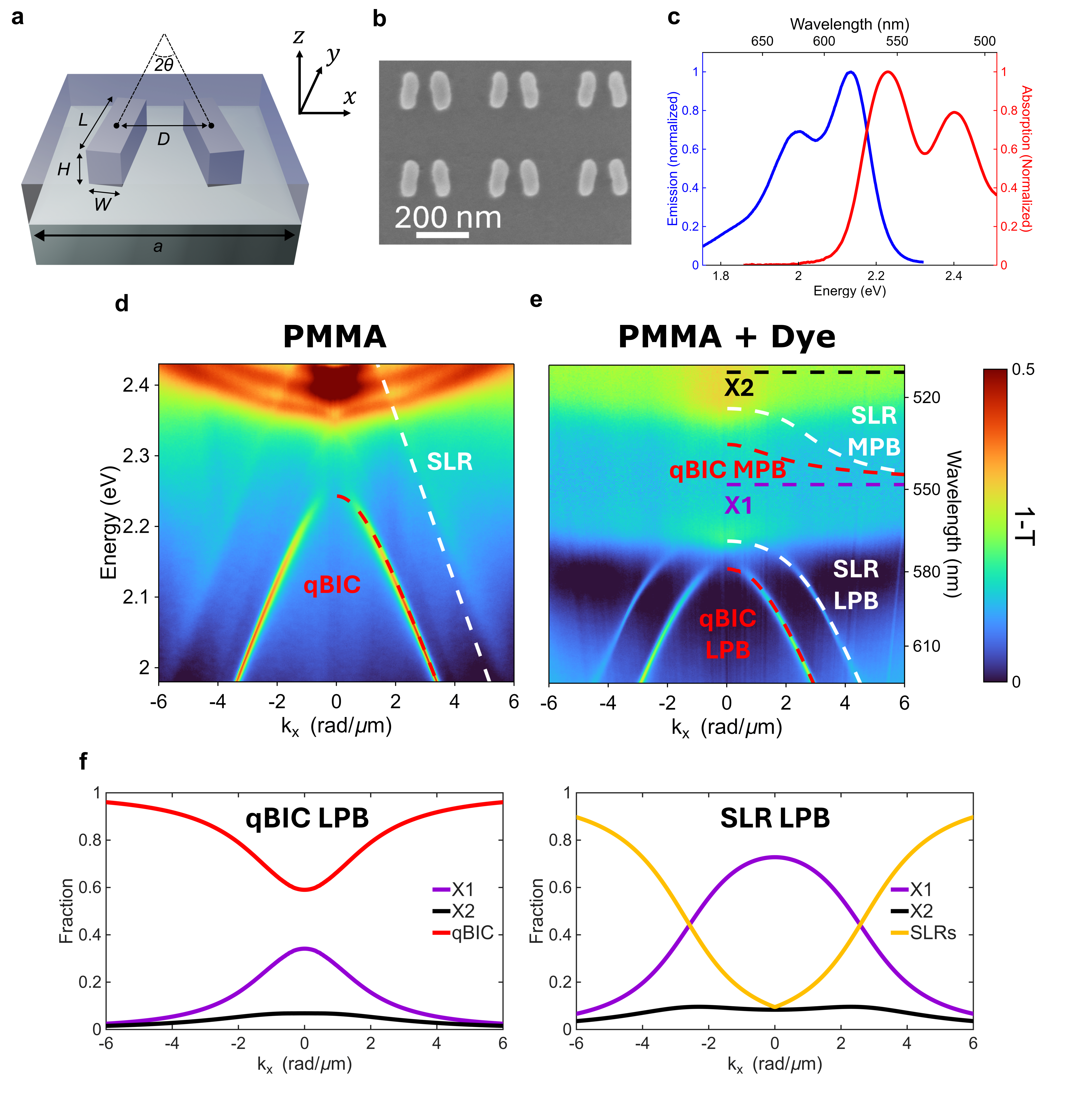}
\caption{\textbf{Metasurface for strong light-matter coupling with excitons in molecular films.} \textbf{a,} Schematic of a single unit cell of the metasurface, consisting of two tilted Si nanorods. \textbf{b,} SEM top view image of the nanorod dimer array on a SiO$_2$ substrate. \textbf{c,} Absorption and photoluminescence spectra of 30wt\% perylene dye in PMMA. \textbf{d,} Extinction map along $k_x$ of the bare Si array coated with a PMMA layer matching the refractive index of the substrate with the incident polarization along the $y$-direction. The SLR and quasi-BIC emerging from the ($\pm1,0$) Rayleigh anomalies are indicated with the white- and red-dashed lines, respectively. \textbf{e,} Extinction map of the Si array coated with the 30wt\% perylene dye in PMMA layer with the incident polarization along $y$-direction. The horizontal dashed lines (X1 and X2) represent the excitonic transitions of the perylene dye. The white and red dashed lines are fits to the dispersion indicating the polariton bands resulting from the coupling of the SLR and quasi-BIC, respectively. Note that only the lower-polariton bands (LPBs) are visible in the measurements. The middle polaritons (MPBs) and upper polaritons (UPBs, not shown) are not visible due to the absorption of the perylene dye.
\textbf{f,} Hopfield coefficients extracted from the fits to the coupled-oscillator model.}
\label{fig:figure_1}
\end{figure}

The optical cavity used for strong light-matter coupling consists of a metasurface of Si nanorods on a SiO$_{2}$ substrate. Details on the fabrication are provided in the Methods section. Each unit cell consists of a pair of tilted Si nanorods with dimensions $H = 90$ nm (height), $W = 50$ nm (width), and $L = 130$ nm (length), separated by a center-to-center distance of $D = 130$ nm, as shown by the schematic in Fig.~\ref{fig:figure_1}a). The dimers are arranged in a square lattice with a periodicity of $a=340$ nm along both in-plane directions, as can be seen in the scanning electron microscope (SEM) image in Fig.~\ref{fig:figure_1}b. This geometry enables the coupling between Mie-type localized resonances (LRs) of the Si nanorods and in-plane diffraction orders or Rayleigh anomalies (RAs), forming surface lattice resonances (SLRs) with high quality factors (Q-factors) in the visible to near-infrared range \cite{Zhu2025}. In addition, a quasi-BIC is formed due to the broken inversion symmetry of the nanorod dimer caused by the tilt of the nanorods by $\theta=10^\circ$. In the case of no tilt between the nanorods, the system has inversion symmetry, and a symmetry-protected bound state in the continuum (BIC) is formed at the $\Gamma-$point (normal direction)~\cite{Plotnik2011,Hsu2016}. The broken in-plane symmetry introduced by the tilt transforms the symmetry-protected BIC into a radiative quasi-BIC, with a large but finite Q-factor~\cite{Koshelev2018}. Quasi-BICs manifest as sharp resonances~\cite{Watanabe2025}, further enhancing the light–matter interaction strength~\cite{KOSHELEV2019}. Although the unit cell defining the metasurface is mirror symmetric and thus has an achiral structure, the array exhibits a strong extrinsic chiral optical response at oblique angles of incidence~\cite{Pura2024}. The tilt angle of $10^{\circ}$ between the two nanorods is specifically chosen as it yields a strong extrinsic chirality in the quasi-BIC resonance. This extrinsic chirality is critical for the control of the spin-selective transport of chiral EPs in directions defined by the dispersion of the array, which is further explained in the following paragraphs. 

As an exciton source, we use a layer (thickness of 470 nm) of the perylene dye at 30 wt\% in PMMA, spin-coated on top of the array. Figure~\ref{fig:figure_1}c shows the absorption and emission spectra of the dye in PMMA. The preparation of the layer is described in the Methods section. The dye possesses an exciton energy at 2.24 eV and a vibronic replica at 2.41 eV. This dye has been investigated in the formation of exciton-polaritons in metasurfaces owing to its robust fluorescence in high-concentration thin films~\cite{Ramezani2017,Castellanos2023,berghuis2024condensation}.

\subsection{Exciton-Polariton Formation}
Angle-resolved extinction measurements are performed using Fourier microscopy (see Methods) to determine whether the optical modes of the metasurface and the excitons in the organic thin film are strongly coupled. The optical extinction, defined as 1-transmittance, is measured for a polarization along the $y$-direction (as defined in Fig.~\ref{fig:figure_1}a), and it is analyzed as a function of the photon energy and the wave vector of the incident beam parallel to the sample surface ($k_{||}$). We focus on the radiation along the $x$-direction ($k_x$), i.e., mainly parallel to the short axis of the nanorods. Figure~\ref{fig:figure_1}d shows the optical extinction measurements as a function of the incident photon energy and the incident wave vector parallel to the surface along the $x$-direction, $k_x$. The extinction maps under different polarization and diffraction directions are shown in the Supplementary Information, Fig.~S2. These measurements correspond to the extinction of the bare array, consisting of the array of Si dimers covered by a 100\% PMMA film (n = 1.49), i.e., an organic film without excitons. The extinction spectra exhibit two distinct bands at low energies with linear dispersions. The higher-energy and weaker band corresponds to the TE-SLR originating from the $(\pm1,0)$ diffraction order RAs, whereas the lower-energy band is assigned to a quasi-BIC arising from the same in-plane diffraction orders~\cite{Pura2024}. 

Figure~\ref{fig:figure_1}e displays the measurements of the strongly coupled sample consisting of a layer of 30 wt\% perylene dye in PMMA on top of the array. The measured transmission is first normalized by the transmission of the thin molecular layer, thereby enhancing the visibility of the polaritons and removing contributions from dark excitons and uncoupled molecular absorption. For consistency, we used the same layer thickness of approximately 470 nm for the bare array (100\% PMMA) and the strongly coupled sample (30 wt\% dye/PMMA). The extinction maps of the strongly coupled system under different polarization and diffraction directions are shown in the Supplementary Information, Fig.~S3. For the strongly coupled sample, the SLR and quasi-BIC spectrally overlap with the absorption band of the dye, giving rise to the formation of the lower-, middle-, and upper-polariton bands (LPBs, MPBs, and UPBs), respectively, and the anti-crossing in the extinction dispersion. The LPBs are visible for $E < 2.2$ eV, while the MPBs and UPBs are not visible due to the overlap with broad molecular absorption of the perylene dye. The LPBs have similar quality factors ($Q\sim55$) to those of the bare photonic modes (SLR and quasi-BIC), indicating that strong exciton–photon coupling preserves the high quality of the underlying modes despite hybridization with the lossy molecular excitons. 

The wave vector difference between the SLR and the quasi-BIC for all frequencies suggests that these photonic modes do not mutually couple to the same electronic transition of a molecule. Therefore, we model the system using two independent ($3\times3$) coupled harmonic oscillator Hamiltonians ($H_\mathrm{qBIC}$ and $H_\mathrm{SLR}$) to quantify the light--matter coupling strengths, where each Hamiltonian comprises one photonic mode (quasi-BIC or SLR) and two excitonic transitions~\cite{Berghuis2023,berghuis2024condensation}. The Hamiltonian of this system is given by 
\begin{equation}\label{coupling hamiltonian}
H_\mathrm{qBIC/SLR}=\begin{pmatrix}
E_\mathrm{qBIC/SLR}(k) - i \gamma_\mathrm{qBIC/SLR}(k) & g_\mathrm{qBIC/SLR}& g_\mathrm{qBIC/SLR}\\
  g_\mathrm{qBIC/SLR} & E_\mathrm{X1}-i\gamma_\mathrm{X1}&0\\
g_\mathrm{qBIC/SLR} &0& E_\mathrm{X2}-i\gamma_\mathrm{X2}
\end{pmatrix},
\end{equation}
where $E_\mathrm{qBIC}(k)$ and $E_\mathrm{SLR}(k)$ are the energy dispersions of the quasi-BIC and SLR modes from the dotted lines in Fig.~\ref{fig:figure_1}d, $E_\mathrm{X1}= 2.26$ eV and $E_\mathrm{X2}=2.42$ eV are the energies of the singlet exciton and the vibronic replica, $\gamma_\mathrm{X1}=128$ meV and  $\gamma_\mathrm{X2}=150$ meV are exciton and vibron dissipation rates obtained from the full-width at half-maximum of a double Gaussian fit to the absorption spectrum of the perylene dye in PMMA~\cite{Berghuis2023,berghuis2024condensation}. $\gamma_\mathrm{qBIC}(k)$ and $\gamma_\mathrm{SLR}(k)$ are the photonic dissipation rates of quasi-BIC and SLRs, respectively, which are obtained from Lorentzian fits for different wave vectors to the extinction measurements in Fig.~\ref{fig:figure_1}d (Supplementary Information, Fig.~S4). $g_\mathrm{qBIC}$ and $g_\mathrm{SLR}$ denote the coupling strengths between the quasi-BIC and excitons and between the SLR and excitons, respectively. The lowest eigen-energies of $H_\mathrm{qBIC}$ and $H_\mathrm{SLR}$ correspond to the lower polariton bands (LPBs). From the coupled oscillator fit to the extinction measurements, indicated with the dashed curves in Fig.~\ref{fig:figure_1}d, we extract the coupling strengths of $g_\mathrm{qBIC}= 97.5$ meV and $g_\mathrm{SLR}= 114$ meV. Comparing to the system dissipation ($\gamma_\mathrm{X1}=128$ meV, $\gamma_\mathrm{X2}=150$ meV and Supplementary Information, Fig.~S4), the system fulfills Savona et al. criterion for strong-light matter coupling, $2g_\mathrm{s} >|\gamma_\mathrm{c}-\gamma_\mathrm{m}|$ where $g_\mathrm{s}$ is the coupling strength and $\gamma_\mathrm{c}$ and $\gamma_\mathrm{m}$ are the loss rates of the cavity (metasurface) modes and the matter (exciton)~\cite{Savona1995}. The weight of the photonic/excitonic character of the polaritons is given by the Hopfield coefficients (Fig.~\ref{fig:figure_1}f), determined by the square of the eigenvector amplitudes in equation~(\ref{coupling hamiltonian})~\cite{Hopfield1958}. These coefficients reveal that either the quasi-BIC or SLR mainly couple to the excitonic transition ($E_\mathrm{X1}=2.26$ eV). The strongest polaritonic hybridization occurs at the $\Gamma$ point for the quasi-BIC, and at $k_x = 3~\mathrm{rad}/\mu\mathrm{m}$ for the SLRs.  

\subsection{Chiral Photoluminescence Dispersion Measurements}
To characterize the chiral emission properties of the LPBs, we measure the circularly polarized photoluminescence enhancement (PLE) dispersion maps. These measurements are done by exciting the sample with a 532 nm continuous-wave laser. A diffuser was placed before the objective lens to depolarize the excitation beam, and the optical power incident on the sample was 2.8 mW. The photoluminescence is measured by placing a quarter-wave plate (QWP) and a linear polarizer (LP) in front of the spectrometer slit to obtain the dispersion for left- and right-circularly polarized emission (see Methods section). The PLE is determined by measuring the emission dispersion of the strongly coupled sample and normalizing it to the emission of the bare dye layer. The PL dispersion maps without normalization are shown in the Supplementary Information, Fig.~S5.

Figures~\ref{fig:dispersion}a and \ref{fig:dispersion}b show the PLE dispersion measurements along $k_x$ for left-handed circularly polarized (LCP) and right-handed circularly polarized (RCP) emission, respectively. The emission is enhanced along the LPBs of the SLR and quasi-BIC, with the LCP emission being more intense for positive in-plane momenta $k_{x}$, and RCP emission dominating for negative $k_{x}$, which reveals the strong extrinsic chiral optical response of the polaritons. At $E = 1.82$ eV, the RCP-PLE reaches a peak value of 7.1 at $k_x \approx -4.5$ rad/$\mu$m, while the LCP-PLE reaches a value of 6.6 at $k_x \approx 4.9$ rad/$\mu$m and $E=1.77$ eV. 

\begin{figure}[!h]
	\centering
	\includegraphics[width=1\textwidth]{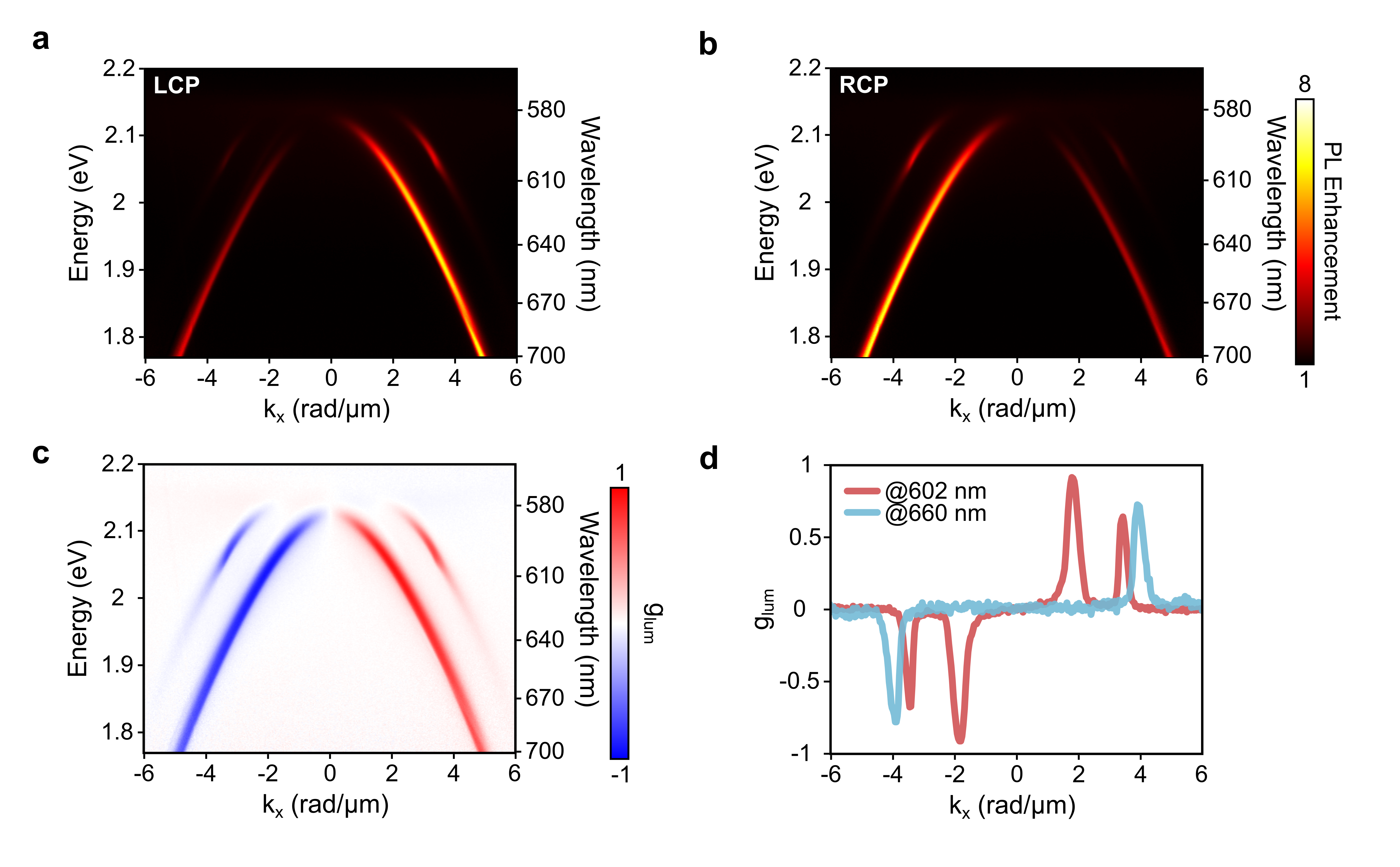} 
    \caption{\textbf{Chiral photoluminescence dispersion of the lower polaritons.} \textbf{a,b,} Angle-resolved photoluminescence dispersion of the left-handed circularly polarized (LCP, \textbf{a}) and right-handed circularly polarized (RCP, \textbf{b}) emission, respectively.  \textbf{c,} Calculated dissymmetry factor $g_{\mathrm{lum}}$ as a function of photon energy and in-plane momentum $k_{x}$. \textbf{d,} Cross-sections of $g_{\mathrm{lum}}$ extracted at 602 nm and 660 nm.}
    \label{fig:dispersion}
\end{figure}

To quantify the dissymmetry of the circularly polarized luminescence (CPL), we determine the $g_\mathrm{lum}$ dispersion map given by
\begin{equation}
g_\mathrm{lum}=2\frac{(I_\mathrm{LCP}-I_\mathrm{RCP})}{(I_\mathrm{LCP}+I_\mathrm{RCP})} \;, 
\end{equation}
where $I_\mathrm{LCP}$ and $I_\mathrm{RCP}$ represent the emission intensities of left-handed and right-handed CPL, respectively. The dissymmetry dispersion is plotted in Fig.~\ref{fig:dispersion}c. The $g_\mathrm{lum}$ dispersion shows the momentum-dependent chiral optical response of exciton-polaritons, with large values $g_\mathrm{lum}$ for $k_{x} \neq 0$. The dissymmetry factor exhibits pronounced extrema at $E = 2.05$ eV, reaching a maximum of $g_\mathrm{lum} \approx 0.93$ and a minimum of $g_\mathrm{lum} \approx -0.92$. Cross-sectional profiles of the dissymmetry factor at 602 and 660 nm are shown in Fig.~\ref{fig:dispersion}d, further highlighting the strong chiral nature of the exciton-polaritons. At 602 nm, two distinct chiral resonances are resolved, corresponding to the quasi-BIC and SLR LPBs. By contrast, at 660 nm, the chiral response is dominated by the quasi-BIC LPB. We note that the integrated $g_\mathrm{lum}$ over all wave vectors at every photon energy is zero, as it is expected from the mirror-symmetric structure of the tilted Si bars. In addition, the SLR polaritons radiating along $k_y$ show a $g_\mathrm{lum}\approx0$ (Supplementary Information, Fig.~S6)

The mechanism leading to the spin-polarized transport of EPs shown in the next section relies on the strong coupling with the extrinsically chiral modes of the metasurface. Strong coupling dresses EPs with chirality, while the extrinsic character of this chirality in the mirror-symmetric metasurface induces the opposite spin for wave vectors with opposite signs, i.e., EPs with opposite spin propagate in opposite directions. This mechanism should not be limited to extrinsically chiral systems; it can also operate in intrinsically chiral metasurfaces, where chiral EPs with a defined spin will propagate as dictated by the dispersion relation. 

The extrinsic chirality of the EPs is further examined by circular dichroism (CD) measurements shown as Supplementary Information in Figs.~S7 and S8. This CD is defined as $\mathrm{CD}=(T_\mathrm{LCP}-T_\mathrm{RCP})/(T_\mathrm{LCP}+T_\mathrm{RCP})$, where $T_\mathrm{LCP}$ and $T_\mathrm{RCP}$ correspond to the transmittance of left- and right-handed circularly polarized light, respectively. The relatively small CD compared with $g_\mathrm{lum}$ arises from different illumination sources, reflecting the distinct physical processes defining the CD and CPL. In addition, extinction dispersion maps under LCP and RCP illumination along $k_y$ show no polarization dependence, and the corresponding CD remains zero (Supplementary Information, Fig.~S8), confirming the achiral optical response along this wave vector and the strong anisotropic response of the sample.

\subsection{Dispersion controlled transport of chiral exciton-polaritons}
\begin{figure}[!h]
	\centering
	\includegraphics[width=1\textwidth]{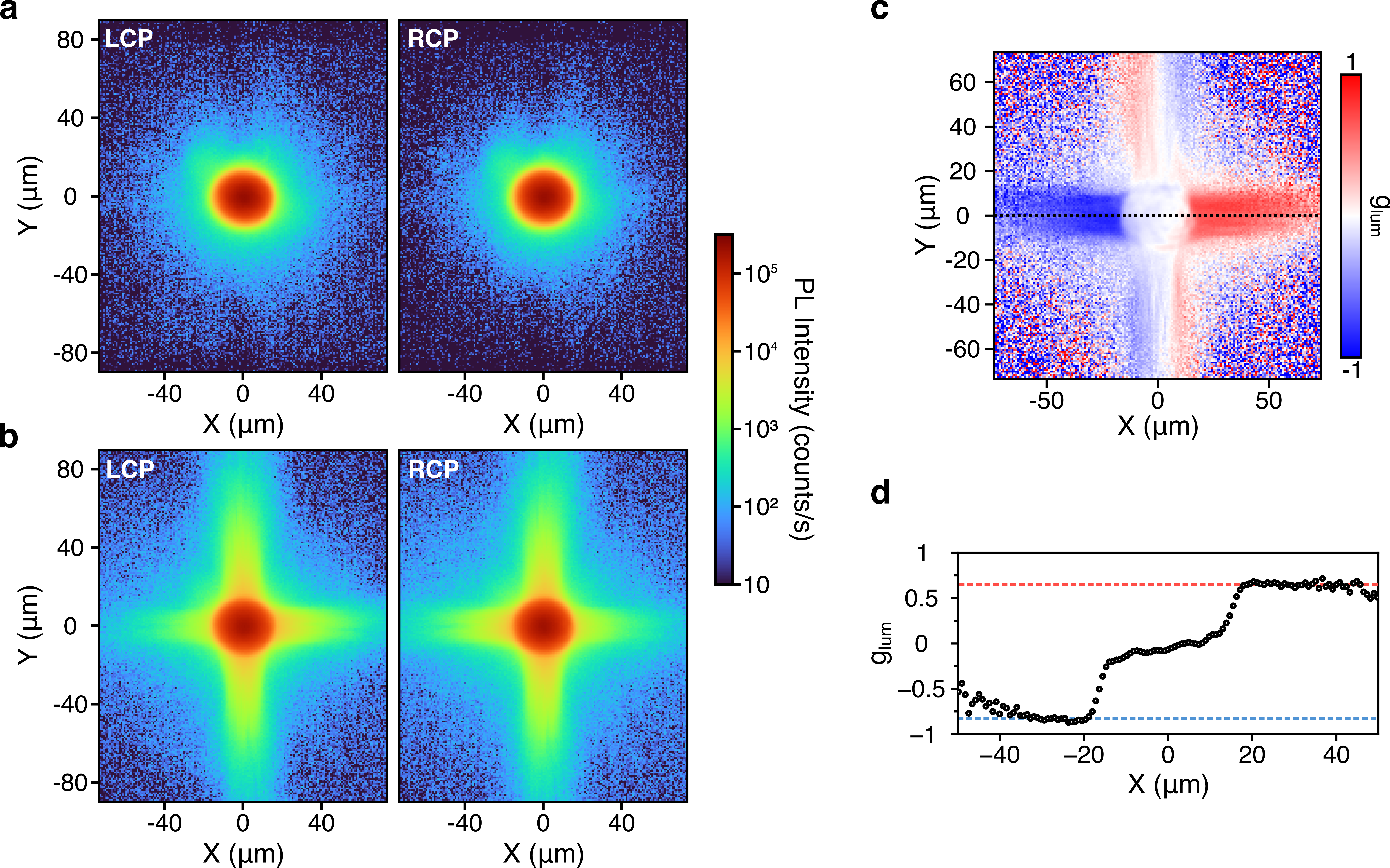}
        \caption{\textbf{Spatially resolved PL and propagation of chiral exciton-polaritons.}  \textbf{a}, Real-space PL images of the bare perylene dye film ($\lambda>550$ nm) for LCP and RCP emission. \textbf{b}, Real-space PL images of dye on top of the Si metasurface for LCP and RCP emission. Emission from EPs is observed far from the excitation spot. This emission occurs in directions defined by the square lattice of Si nanorods. \textbf{c}, Real space image of the dissymmetry factor ($g_{\mathrm{lum}}$) of the strongly coupled system, showing the directional propagation of chiral EPs. \textbf{d}, Cross-section of the dissymmetry factor along the $x$-direction at $y$ = 0~$\mu$m (along the dashed line in \textbf{c}). The dissymmetry of chiral EPs remains larger than 0.6 over distances as long as 40 $\mu$m, as indicated by the horizontal dashed lines.}
    \label{fig:real_space} 
\end{figure}

To investigate the transport of chiral EPs, we perform PL imaging measurements in the Fourier microscope by adding an additional lens in front of the camera to capture the real-space image of the polariton emission. One pinhole is added before the excitation to control the laser spot size. Figure~\ref{fig:real_space}a shows the emission of the bare dye layer for LCP (left panel) and RCP (right panel). Emission is only observed from the area that is illuminated with the excitation laser, as expected from the short propagation distance of Frenkel excitons in organic materials of only a few tens of nanometers~\cite{Mikhnenko2015}, i.e., well below the diffraction limit of our optical microscope. In contrast, in the strongly coupled system (Fig.~\ref{fig:real_space}b), we observe directional polaritonic propagation over several tens of micrometers along both the $x$- and $y$-directions, for both, LCP (left panel) and RCP (right panel) emission. The resulting cross pattern emerges from the square lattice of Si nanorods. 

To analyze the chiral optical properties of the coupled system, we construct the spatially resolved $g_{\mathrm{lum}}$ maps from the LCP and RCP emission for wavelengths longer than 550 nm. As expected, the bare dye film measurement (Supplementary Information, Fig.~S9) shows that the $g_{\mathrm{lum}}$ remains zero across the emission profile, consistent with the achiral nature of the molecules. In contrast, the strongly coupled system exhibits a significant propagation of chiral EPs mostly along the $x$-axis (Fig.~\ref{fig:real_space}c). This result agrees with the PLE dispersion analysis, where LCP emission dominates for positive in-plane momentum $k_{x}$ (propagation along $x>0$), while RCP emission dominates for negative $k_{x}$ (propagation along $x<0$). In addition, $g_{\mathrm{lum}}\approx0$ along the $y$-axis, which is consistent with the achiral optical response in the dispersion map for $k_y$, as shown in the Supplementary Information, Fig.~S6. We note that the central spot, showing $g_{\mathrm{lum}}\approx0$, is caused by uncoupled and dark excitons from the reservoir where they are generated~\cite{Dang2024}. According to the cross-section of the spatial map along the $x$-direction shown in Fig.~\ref{fig:real_space}d, $g_\mathrm{lum}$ reaches a maximum of 0.71 at $x=37~\mu$m along the $x>0$ direction and a minimum of –0.87 at $x=-24~\mu$m along the $x<0$ direction from the excitation center. These remarkably high values of $g_{\mathrm{lum}}$ remain constant up to distances of $\sim40$ $\mu$m from the center before gradually decaying along with the decay of the exciton-polariton density far from the reservoir.

\begin{figure}[!h]
	\centering
	\includegraphics[width=0.9\textwidth]{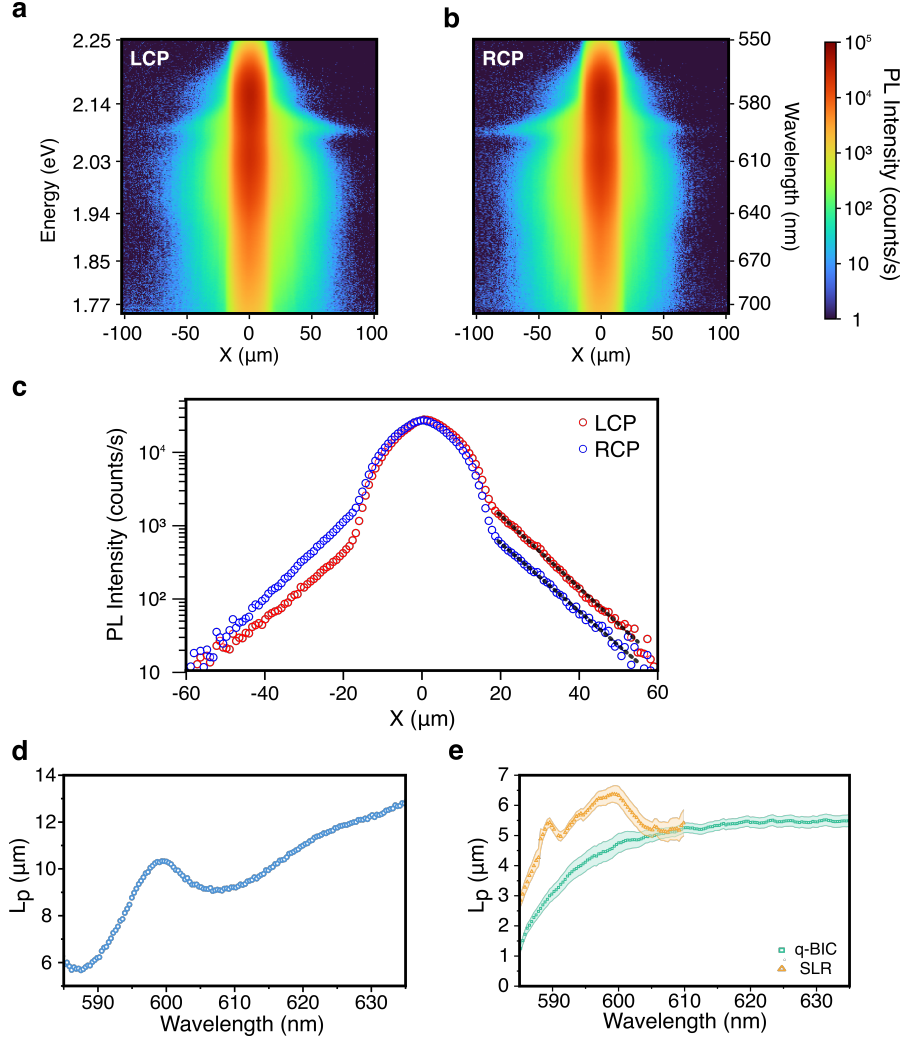}
        \caption{\textbf{Energy-resolved propagation of chiral exciton-polaritons.} Real-space intensity maps of EPs as a function of energy and distance along the $x$-direction from the excitation center for \textbf{a} LCP and \textbf{b} RCP emission. \textbf{c}, Cross-section of emission intensity at 610 nm along the $x$-direction for LCP and RCP. The dashed lines in the positive $x$ side represent the exponential fits to obtain the propagation length. \textbf{d}, Wavelength-dependent propagation length extracted from \textbf{a} and \textbf{b}. \textbf{e}, Wavelength-dependent propagation length extracted from the dispersion measurements (Fig.~\ref{fig:dispersion}a and \ref{fig:dispersion}b).} 
    \label{fig:propagation}
\end{figure}

Energy-resolved propagation maps of the chiral EPs are obtained by closing the spectrometer slit to spectrally disperse the emitted light (see Methods). The maps for LCP and RCP emissions are shown in Figs.~\ref{fig:propagation}a and \ref{fig:propagation}b, respectively. In agreement with the momentum-space and real-space measurements (Figs.~\ref{fig:dispersion} and \ref{fig:real_space}b), LCP emission predominates for positive distances, whereas RCP emission is more prominent for negative distances. The spatial dependence of the emission, defining the propagation of chiral EPs, is further illustrated in Fig.~\ref{fig:propagation}c. 
The bright central region at $|x|<15~\mu$m corresponds to emission from the exciton reservoir and has equal intensity of LCP and RCP emission, as expected from the non-chiral dye molecules (see Supplementary Information, Fig.~S10). The regions at longer distances ($|x|>15~\mu$m) exhibit pronounced helicity-dependent emission associated with the propagation of chiral  EPs. In addition, the propagation characteristics of the EPs display a clear energy dependence owing to their dispersive nature. To quantitatively analyze the propagation, we fit the spatial profiles for $|x|>15~\mu$m using a single-exponential decay, as shown by the dashed lines in Fig.~\ref{fig:propagation}c for $x>0$. The exponential decay constant defines the EP propagation length, denoted as $L_p$. We note that the propagation lengths obtained from the LCP and RCP emission are identical, as both originate from the same LPB. The dissymmetry of the chiral EP emission manifests through the different intensities of LCP and RCP characterized by the pre-factors of the exponential fits, which remain independent of the distance (Fig.~\ref{fig:propagation}c). These observations further validate the robust propagation of chiral EPs over long distances, preserving a well-defined polarization state.

The wavelength dependence of the fitted $L_p$ is summarized in Fig.~\ref{fig:propagation}d. The $L_p$ increases with increasing wavelength, reflecting the progressively stronger photonic character of the EPs as described by the Hopfield coefficients (see Fig.~\ref{fig:figure_1}f). Intriguingly, a local maximum with $L_p>10~\mu$m occurs around 600 nm, coinciding with the spectral region where both the SLR and quasi-BIC LPBs co-exist (See Figs.~\ref{fig:dispersion}a and \ref{fig:dispersion}b). To resolve the contribution of the two EPs, we estimate the $L_p$ for each wavelength from the far-field emission as the inverse of the imaginary component of the wave vector of the LPB, corresponding to $1/\Delta k_x$ \textcolor{red}{\cite{Berghuis2022}}. $\Delta k_x$ was determined by fitting the far-field emission spectra of Fig.~\ref{fig:dispersion} for each wavelength with a Lorentzian. This approach allows the total propagation length to be decomposed into the contributions from the chiral EPs emerging from the strong coupling to the SLR and quasi-BIC. As shown in Fig.~\ref{fig:propagation}e, the local maximum of the $L_p$ at 600 nm can be attributed primarily to the propagation of the chiral EPs originating from the SLR. The narrower momentum linewidth ($\Delta k_x$) of these EPs leads to longer propagation lengths than those originating from the quasi-BIC. This result is in line with the different group velocities of the EPs (see Supplementary Information, Fig.~S11). In particular, the group velocity is larger for the EPs originating from the SLR. We note that the $L_p$ extracted from the real-space measurements (Fig.~\ref{fig:propagation}d) are consistent with those obtained from the momentum-space analysis (Fig.~\ref{fig:propagation}e). Specifically, the sum of the propagation lengths associated with the SLR and quasi-BIC chiral EPs reproduces the total propagation length measured in real space, further validating the interpretation of the distinct chiral EPs contributions.

\begin{figure}[h]
	\centering
	\includegraphics[width=0.5\textwidth]{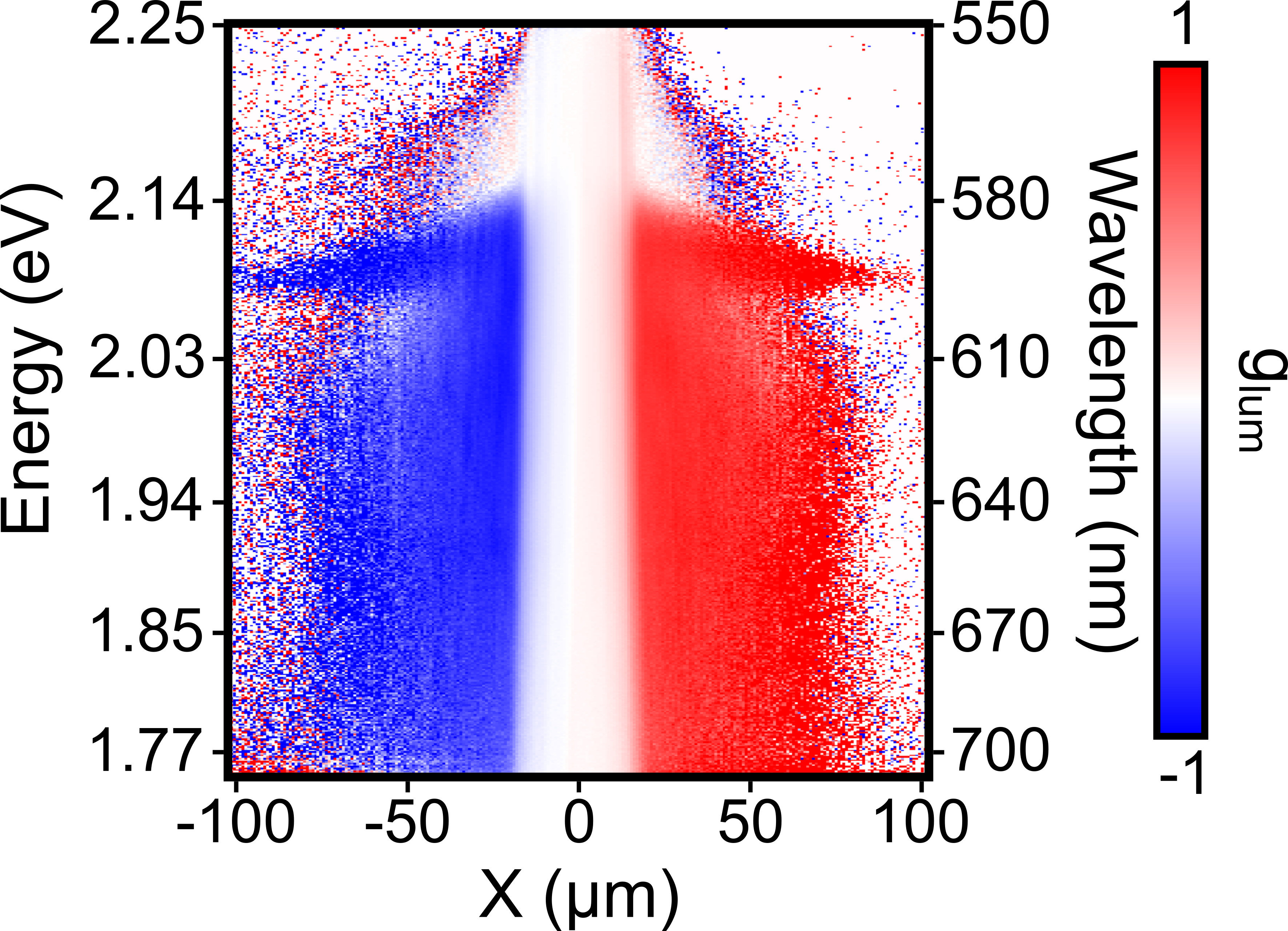}
        \caption{\textbf{Energy-resolved propagation of the dissymmetry from chiral exciton-polaritons.}  Real-space $g_{\mathrm{lum}}$ maps of chiral EPs as a function of energy and distance along the $x$-axis from the excitation center spot at $x$ = 0.}
    \label{fig:g factor map}
\end{figure}

Based on Fig.~\ref{fig:propagation}a and \ref{fig:propagation}b, we further construct the energy-resolved dissymmetry propagation map shown in Fig.~\ref{fig:g factor map}. These results reveal that EPs at different energies exhibit similar dissymmetry factors with $|g_{\mathrm{lum}}| \sim 0.9$, which remain nearly constant over long propagation distances. This observation is consistent with the real-space measurements shown in Fig.~\ref{fig:real_space}, indicating that the spin-angular momentum encoded in the chiral EPs is preserved during propagation through the strong light-matter coupling. In addition, we expect that the extended $L_p$ due to the SLR EPs also leads to a longer propagation of $g_{\mathrm{lum}}$. Either the quasi-BIC EPs or the SLR EPs have a nearly constant $g_{\mathrm{lum}}$ over 50 $\mu$m. The decay of the $g_{\mathrm{lum}}$ is associated with the decay of the EPs, suggesting that the preservation of dissymmetry is governed by the $L_p$. These results demonstrate the unique capability of strongly coupled systems to enable long-range transport of spin-polarized EPs, while maintaining a high degree of optical chirality.


\section*{Conclusions}
In summary, we have demonstrated enhanced transport of chiral exciton-polaritons by using a dielectric metasurface as a cavity for strong light-matter coupling. Extrinsically chiral SLRs and quasi-BICs supported by the metasurface strongly couple to excitons in an organic film, leading to the formation of chiral EPs with high dissymmetry factors, $g_{\mathrm{lum}}\sim0.93$, and defined spin that propagate in directions given by the dispersion of the metasurface. This approach is robust, highly tunable, and compatible with a wide range of ambient-stable excitonic materials. The real-space $g_{\mathrm{lum}}$ images, together with the propagation length analysis, show spin-polarized transport with a constant $g_{\mathrm{lum}}$ over distances of 50 $\mu$m. These results open a new direction in the field of polaritonics. The robust spin-dependent transport of chiral EPs demonstrated here has broad implications for quantum and optoelectronic applications. Chiral EPs offer a natural route toward compact on-chip sources of spin-polarized transport, including low-threshold spin-selective polariton lasers~\cite{Hong2025,Wang2023}, and CISS-induced spin polarization~\cite{bloom2024chiral,naaman2012chiral}. In quantum information processing, the ability to route and store information in spin-polarized polariton modes suggests new approaches to optical logic gates, and quantum networks operating at ambient conditions~\cite{Opala2023,Chen2021}. 

\section{Methods}\label{Methods}
\subsection{Sample Preparation}
The perylene dye is provided by Dr. Martin K\"{o}nemann. Mixed films containing 30 wt\% perylene dye (N,N$'$-bis(2,6-diisopropylphenyl)-1,7- and 1,6-bis(2,6-diisopropylphenoxy)-perylene-3,4:9,10-tetracarboximide) and PMMA (Sigma-Aldrich) were prepared from toluene and chloroform (1:2 v/v) solutions, with solute concentrations of 12 mg/mL and 28 mg/mL, respectively. The solutions were spin-coated at 1000 rpm, yielding films with a thickness of approximately 470 nm, as measured using a Dektak profilometer.

\subsection{Fabrication of Si metasurfaces}
Polycrystalline Si thin films (90 nm) were deposited on silica glass substrates by low-pressure chemical vapor deposition using SiH$_4$ as the precursor gas. A positive resist (ZEP520A) was spin-coated onto the Si layer and patterned by electron-beam lithography, yielding nanorod hole arrays in the resist after development. A 70 nm Cr layer was then deposited by electron-beam evaporation. Lift-off defined Cr nanorod arrays on the Si surface. These Cr structures served as an etch mask for vertical dry etching of the underlying Si film, performed using the Bosch process with SF$_6$ and C$_4$F$_8$ gases. Finally, the Cr mask was removed by wet etching in an acidic solution (S-clean S24, Sasaki Chemical Co., Ltd.). The fabricated Si nanorod arrays covered an area of 2~$\times$~2 mm$^2$.

\subsection{Fourier microscopy}
The setup used for the measurements has been reported in previous studies~\cite{liang2024tailoring,Zhu2025}. The extinction and PL dispersion maps of the SLRs, quasi-BIC, and polariton bands were measured using a Fourier microscope. The sample was illuminated through a 40 $\times$ objective (Nikon CFI S Plan Fluor ELWD, NA = 0.6), and the transmission and PL were collected with a 60 $\times$ objective (Nikon CFI S Plan Fluor ELWD, NA = 0.7). Angle-resolved spectra were recorded with an imaging Princeton Instruments SP2300 spectrometer coupled to a ProEM:512 camera (Princeton Instruments), enabling the mapping of the dispersion as a function of energy and emission angle. Photoluminescence excitation was provided by a continuous-wave laser at 532 nm with an incident power of 2.8 mW. Real-space PL images were obtained by inserting an additional lens to project the real-space image, instead of the Fourier image, onto the detection plane.

\backmatter
\bibliography{sn-bibliography}


\bmhead{Supplementary information}
Supplementary Figs.~1-10.

\bmhead{Acknowledgements}
This project was funded by the European Union. Views and opinions expressed are however those of the author(s) only and do not necessarily reflect those of the European Union or the European Innovation Council and SMEs Executive Agency (EISMEA). Neither the European Union nor the granting authority can be held responsible for them. (SCOLED, Grant Agreement No. 101098813) Y.-C.W. acknowledges the financial support from the National Science and Technology Council (NSTC) (115-2113-M-007-014-MY3; 114-2639-M-002-008-ASP). S.M. acknowledges financial support from JSPS, Japan (JP25K01501, JP25K21709, JP26H02198). We thank Jacub Grečner for the assistance with the circular dichroism measurements.

\section*{Declarations}
Not applicable




\end{document}